\documentstyle [12pt,epsf] {article} 
\begin{document}
\begin{titlepage}
\begin{center}
\vspace{2cm}
\LARGE
Clustering of Galaxies in a Hierarchical Universe:  II. Evolution to High Redshift    
\\                                                     
\vspace{1cm} 
\large
Guinevere Kauffmann, J\"{o}rg M. Colberg, Antonaldo Diaferio \& Simon D.M. White  \\
\vspace{0.5cm}
\small
{\em Max-Planck Institut f\"{u}r Astrophysik, D-85740 Garching, Germany} \\
\vspace{0.8cm}
\end{center}
\normalsize
\begin {abstract}
In  hierarchical cosmologies the evolution of galaxy clustering depends both on
cosmological quantities such as $\Omega$, $\Lambda$ and $P(k)$, which determine
how collapsed structures --  dark matter halos -- form and evolve, and
on the physical processes -- cooling, star formation, radiative and
hydrodynamic feedback -- which drive the formation of 
galaxies within these merging halos.  In this paper, we combine 
dissipationless cosmological N-body simulations and semi-analytic models of 
galaxy formation in order to study how these two aspects interact.
We focus on the differences in clustering predicted for galaxies of
differing luminosity, colour, morphology and star formation rate, and
on what these differences can teach us about the galaxy formation process.
We show that a ``dip'' in the amplitude of galaxy correlations between
$z=0$ and $z=1$ can be an important diagnostic. Such a dip occurs
in low-density CDM models, because structure forms early and dark matter
halos of mass $\sim 10^{12} M_{\odot}$, containing galaxies with luminosities $\sim L_*$,  
are unbiased tracers of the dark matter over this redshift range;
their clustering amplitude then evolves similarly to that of the dark
matter. At higher redshifts bright galaxies become strongly biased and the
clustering amplitude increases again. In high density models, structure forms late
and bias evolves much more rapidly.
As a result, the clustering amplitude of $L_*$ galaxies
remains constant from $z=0$ to $z=1$. The strength of these effects
is sensitive to sample selection. The dip becomes weaker for galaxies
with lower star formation rates, redder colours, higher luminosities
and earlier morphological types. We explain why this is the case and
how it is related to the variation with redshift of the abundance and
environment of the observed galaxies. We also show that the relative
peculiar velocities of galaxies are biased low in our models, but that
this effect is never very strong.  Studies of clustering evolution
as a function of galaxy properties should place strong constraints on
models of galaxy formation and evolution.

\end {abstract}
\vspace {0.8cm}
\end {titlepage}

\section {Introduction}          
Local galaxies are highly clustered. On large scales they are
organized into a network of
sheets and filaments which surround large underdense regions, usually
referred to as voids. On smaller scales galaxies are found
in gravitationally-bound groups and clusters. 
According to the standard theoretical paradigm, the structures observed today were formed by the
gravitational amplification of small perturbations in an initially
gaussian dark matter density field. Small scale overdensities were  
the first to collapse, and the resulting objects subsequently merged
under the influence of gravity to form larger structures such as clusters and superclusters.
Galaxies formed within dense {\em halos} of dark matter, where gas was able to reach high
enough overdensities to cool, condense and form stars.

In this hierarchical formation picture, the clustering of the 
dark matter, as measured by the amplitude of the
matter correlation function $\xi_m (r)$, increases monotonically
with time. The precise evolution of $\xi_m (r)$ with redshift has been studied extensively using both
N-body simulations (e.g. Jenkins et al. 1998) and analytic methods ( Hamilton et al 1991; 
Peacock \& Dodds 1994; Jain, Mo \& White 1995).
If $\xi_m(r,z)$ were observable, it would be straightforward 
to use its behaviour to determine $\Omega$, $\Lambda$
and the power spectrum of linear density fluctuations. What one measures
in practice, however, is the clustering of galaxies, and the
interpretation then requires
an understanding of how these objects trace the underlying dark matter density field.

If galaxies form at the centre of dark matter halos, considerable
insight may be gained by using N-body simulations to study the
clustering of halos (Brainerd \& Villumsen 1994; Mo \& White 1996; 
Roukema et al 1997; Jing \& Suto 1998;  Wechsler et al 1998; Bagla 
1998a,b; Ma 1998). Mo \& White (1996) tested an approximate analytic
theory against their numerical results, and this theory and its extensions
can also be used to analyse the evolution of halo clustering with
redshift (Matarrese et al 1997; Coles et al 1998).
An important conclusion from all these studies is that the clustering 
of halos of galactic mass ($\sim 10^{12} M_{\odot}$) evolves much more
slowly than the clustering of the dark matter.
This is because at high redshifts, such halos correspond to rare peaks in
the initial density field, and are thus more strongly clustered than
the dark 
matter (Kaiser 1984). Another important conclusion is that more massive halos
are more strongly clustered than less massive halos. If the luminosity of a galaxy is correlated
with the mass of its halo, more luminous galaxies ought to be more strongly clustered.
A detailed comparison with observational data 
requires a model for the observable properties of the galaxies present
within halos of given mass at each  epoch.                                    

Precise measurement of the clustering amplitude of galaxies at high redshift has just recently become
feasible. Usually this is done by calculating the angular two-point correlation function $w(\theta)$
as a function of apparent magnitude.  In order to assess how clustering
has evolved, $w(\theta)$ must be deprojected using Limber's equation
under the assumption of some specific model for the redshift
distribution of the observed galaxies. In future large surveys of faint galaxies
with photometric and/or spectroscopic redshifts will be available
(see, for example, Connolly et al 1995). It will be possible 
to classify the galaxies in these surveys according to absolute magnitude, 
spectral type, star formation rate and colour,
and to investigate how clustering evolution depends on these properties.

At present, most of the data indicate that the clustering amplitude of galaxies
{\em decreases} from $z=0$ to $z=1$. Different analyses, however, yield
very different estimates for the strength of this decrease. Le F\`{e}vre et al (1996) analyzed the
clustering of 591 galaxies with $I<22.5$ in the five 10 arcminute
fields of the CFRS survey. They find
that clustering has evolved dramatically, quoting a comoving correlation length
at redshift 0.5 of $r_0= 2$ h$^{-1}$ Mpc ($q_0=0.5$). 
More recently, Carlberg et al (1998) presented a preliminary analysis 
of clustering in the CNOC2 field galaxy redshift survey. Their sample is spread over four
patches of sky with a total area of 1.5 square degrees. They estimate
that the comoving correlation length of galaxies with
$M_R < -20$ evolves as $r_0(z)= (5.15 \pm 0.15) (1+z)^{-0.3 \pm 0.2}$ h$^{-1}$ Mpc.
This is much weaker than the evolution found for the CFRS galaxies. The results of Carlberg et al (1998) 
agree reasonably well with those of Postman et al (1998), based on 710 000 galaxies
with $I_{AB} < 24$ from an imaging survey of a contiguous 4 square degree region of the sky.
The latter authors suggest that the small volumes sampled by the CFRS and other early surveys
resulted in their derived correlation lengths being biased low.

The clustering of Lyman break galaxies at $z \sim 3$ has now been
measured with surprising precision
(Steidel et al 1998a; Giavalisco et al 1998; Adelberger et al 1998). The comoving correlation length
of these objects is comparable to that of $L_*$ galaxies today,
implying, as expected from models of halo clustering, that Lyman 
break galaxies are highly biased tracers of 
the dark matter distribution at these redshifts. In addition,
Giavalisco et al (1998)
find that the fainter Lyman break galaxies are less strongly
clustered. This accords well with a simple model in which the star 
formation rates in these objects increase with the mass of their halos. More detailed theoretical modelling
of the observed properties of Lyman break galaxies at $z=3$, including
analysis of their abundances, sizes,
luminosities, colours, star formation rates and clustering properties,
has been carried out by Mo \& Fukugita (1996), 
Baugh et al (1998), Governato et al (1998), Somerville, Primack \& Faber(1998) and Mo, Mao \& White (1998)

In this paper, we combine cosmological N-body simulations and
semi-analytic modelling
of galaxy formation to study the evolution of  galaxy clustering as a function of redshift.
Our methods for incorporating galaxy formation in the simulations are 
discussed in detail in Kauffmann et al (1998, Paper I). Two variants
of a cold dark matter (CDM) cosmology are
analyzed here: a high-density model with $\Omega=1$, $\Gamma=0.2$ and $H_0 = 50$ km s$^{-}$ Mpc$^{-1}$
($\tau$CDM), and a low-density flat model with $\Omega=0.3$, $\Lambda=0.7$ and 
$H_0= 70$ km s$^{-1}$ Mpc$^{-1}$ ($\Lambda$CDM). Paper I was concerned with the global properties of the
galaxy distribution at $z=0$, including  B and K-band luminosity functions,
the I-band Tully-Fisher relation, galaxy two-point correlation functions,  colour
distributions, star formation rate functions and peculiar velocity distributions.
Here we focus on clustering evolution  in the two models.
We study the predicted differences in clustering evolution
for galaxies of different magnitude, type and star formation rate, and we outline            
how future observational data will clarify the galaxy formation process.
 
 \section {What can be learned from the evolution of halo clustering?}
Because galaxy formation is complex and involves many
poorly-understood physical processes, for example, star formation and
radiative and hydrodynamical feedback, it is worthwhile to ask whether the
clustering of dark matter halos can be used to constrain cosmological
parameters directly.

In figure 1 we plot the correlation length  $r_0$ as a function of redshift for halos of different
mass in our two simulations. Here, as in the rest of the paper, all length scales are
expressed in comoving units.
Since the correlation functions in the models are not exact power laws, we 
define $r_0$ as the radius where $\xi(r)=1$. The smallest halos resolved in the simulations
contain 10 particles and have virial masses $\sim 2\times 10^{11} M_{\odot}$. 
For these objects, $r_0$ initially decreases with redshift,
reaches a minimum, and then increases again. The redshift of this
minimum is different for the two cosmologies: $z \sim 0.7$ for $\tau$CDM and
$z\sim 1.5$ for $\Lambda$CDM. Massive halos do not exhibit the same ``dip'' in 
correlation length; their $r_0$ remains constant for a while, then increases at high redshift.
Once again, the redshift at which the evolution becomes strong
is lower for $\tau$CDM than for $\Lambda$CDM.
This is simply because structure formation occurs {\em later} in the $\tau$CDM
model.

\begin{figure}
\centerline{
\epsfxsize=10cm \epsfbox{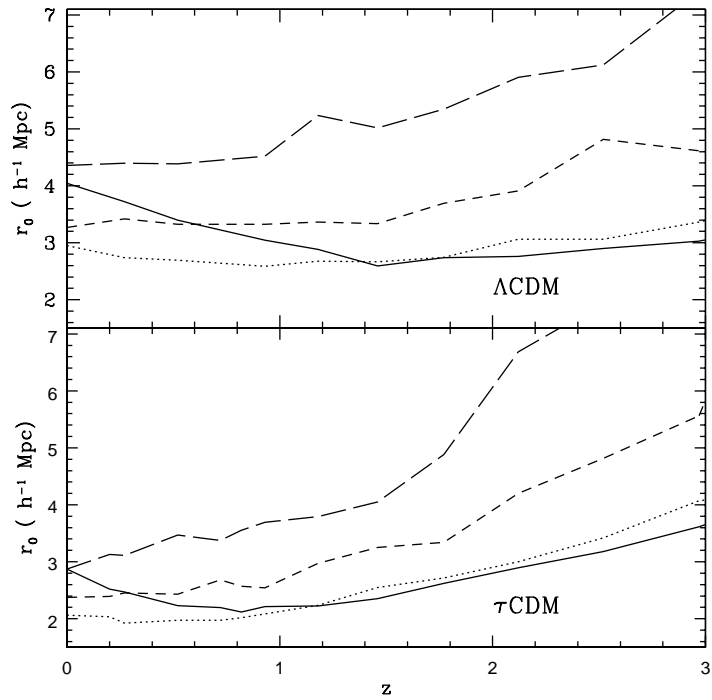}
}
\caption{\label{fig1}
The evolution of the co-moving correlation length of halos as a function of redshift in
the $\tau$CDM and $\Lambda$CDM simulations. The solid line is for halos with $\log (M_{vir}/M_{\odot})$
in the range $11.0-11.5$, the dotted line for $11.5-12$, the
short-dashed line for $12-12.5$ and the long-dashed  
line for $12.5-13.$}
\end {figure}

With 10 metre telescopes, it is now possible to measure the rotation curves of disk galaxies
at redshift $\sim 1$ (Vogt et al 1996). It will 
be many years, however, before such samples are both  large enough  and complete enough for 
an analysis of the clustering
evolution of galaxies as a function of their halo mass. In all likelihood, we will have to 
deal with flux-limited
surveys of galaxies for some time to come.             

Let us now make the simplifying assumption that each simulated dark
matter halo contains one observable galaxy, and that the luminosity of
the galaxy increases with the mass of its halo.
The correlation function of a flux-limited sample of galaxies {\em of known abundance} at redshift $z$     
may then be calculated by evaluating $\xi(r)$ for the mass-limited set
of simulated halos which has the same abundance.
This is illustrated in figure 2, where we plot the correlation length of halos versus
their number density (in units of h$^{3}$ Mpc$^{-3}$) at a series of redshifts.
As expected, $r_0$ decreases as the
number density increases, because the correlation signal becomes dominated by
low-mass halos, which are more weakly clustered. Note that the differences between the $\tau$CDM
and the $\Lambda$CDM models are  small at all redshifts. Mo, Mao \& White (1998)
show a similar plot for halos at $z= 3$ for four different CDM
cosmologies and find that they all
give similar results. There thus appears to be  a ``cosmic conspiracy'' that makes it
impossible to infer information about cosmological parameters from the clustering
of halos of given abundance. On the other hand, the uniformity seen in
figure 2 can be used as a {\em test} 
of the entire class of hierarchical models and of the hypothesis that there is a one-to-one correspondence
between halo mass and galaxy luminosity. As discussed by Steidel et al (1998b) and by
Mo, Mao \& White (1998), this hypothesis works well for the Lyman break
population at $z \sim 3$.
At low redshifts, the assumption of a one-to-one correspondence between halos and galaxies
must break down, because the abundance of high mass halos is larger, and more and more
bright galaxies are grouped together in each such halo. The values of $r_0$ plotted in figure
2 are then lower limits on the true values.

\begin{figure}
\centerline{
\epsfxsize=11cm \epsfbox{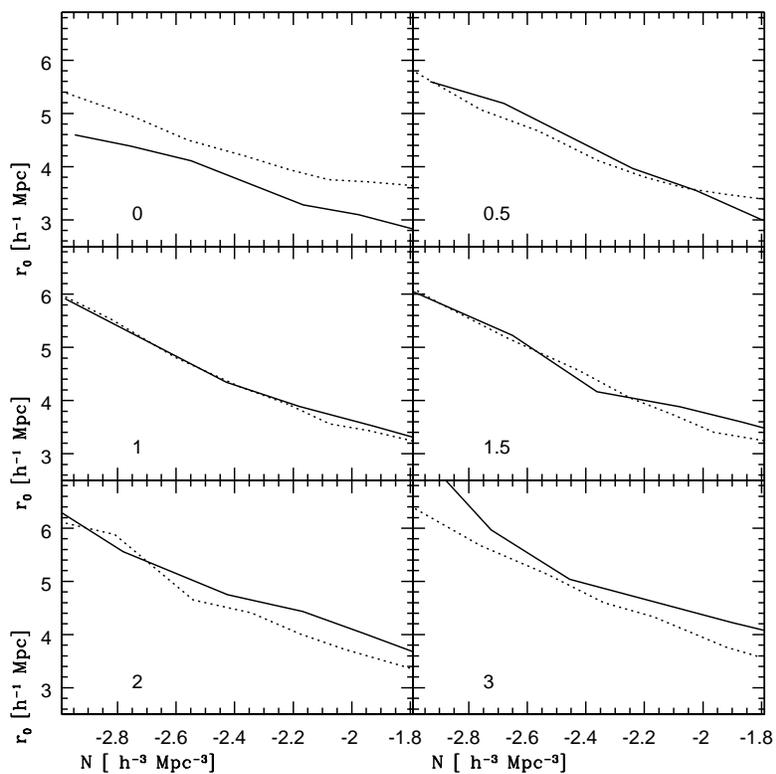}
}
\caption{\label{fig2}
The co-moving correlation length $r_0$ of halos is plotted against comoving number density
at redshifts  0, 0.5, 1, 1.5, 2  and 3. The solid line shows results for the $\tau$CDM
simulation and the dotted line for the $\Lambda$CDM simulation.}
\end {figure}
    
\section {The evolution of galaxy clustering}
In this section, we study the evolution of galaxy clustering in the
$\Lambda$CDM and $\tau$CDM simulations. We use the star formation and feedback recipes that resulted in the
best fits to the observational data at $z=0$. As discussed in Paper I, extremely efficient 
feedback was required in the $\tau$CDM model in order to obtain a reasonable fit to the
correlation function on scales below 1 h$^{-1}$ Mpc and to avoid producing too many 
galaxies with luminosities below $ L_*$. Even so, the model failed to
fit the observed bright end of the  luminosity function  and the clustering amplitude
was too low on large scales. The $\Lambda$CDM model with relatively inefficient feedback 
resulted in a better
overall fit to most of the data at $z=0$. 
For simplicity, we do not consider dust extinction in the analysis of
this paper because it is very uncertain how the empirical recipes we adopted
in Paper I should be extended to high redshift. This neglect has little effect
on our $\Lambda$CDM model but means that our $\tau$CDM model now
substantially underpredicts galaxy clustering at $z=0$. We concentrate below on
the relative evolution of $r_0$ rather than on its absolute value, so
this problem does not strongly affect our conclusions.

In figure 3 results are shown for galaxies with rest-frame B-band magnitudes brighter than
$-19 +5 \log h$ in the $\Lambda$CDM simulation.
At redshift zero, this corresponds to selecting galaxies brighter than $\sim L_*$. 
The first three panels in the plot show the evolution of $\xi(r)$ evaluated 
 at $r=$ 2, 3 and 8 h$^{-1}$ Mpc (co-moving units). The fourth panel shows the evolution
of the co-moving correlation length $r_0$. For comparison, the dotted line in each panel 
shows the evolution of the corresponding quantity for the dark matter.
Results for $\tau$CDM are given  in figure 4. In each case the
redshift extends to the point at which the abundance of $L_*$
galaxies
becomes too low for reliable estimation of the correlation function.

In the $\Lambda$CDM model, the clustering amplitude decreases from $z=0$ to $z=1.5$, remains
approximately constant from $z=1.5$ to $z=2.5$ and then increases again at higher redshift.
The dip in clustering amplitude is stronger on small scales: $\xi(r)$ decreases by a factor
 of 3 at 2 h$^{-1}$ Mpc and by a factor 
1.5 at 8 h$^{-1}$ Mpc. The correlation length $r_0$ decreases from
5.5 h$^{-1}$ Mpc at $z=0$ to 3.9 h$^{-1}$ Mpc at $z=1.5$. This agrees
remarkably well with the parametrization of $r_0$ as a function of $z$ quoted
by Carlberg et al (1998). In the $\tau$CDM model the clustering amplitude remains fixed  
from $z=0$ to $z = 1$ and then rises steeply at higher
redshifts. Note, as mentioned above, that the
correlation length at $z=0$ is low ($\sim 3 $ h$^{-1}$ Mpc) in this model.

\begin{figure}
\centerline{
\epsfxsize=14cm \epsfbox{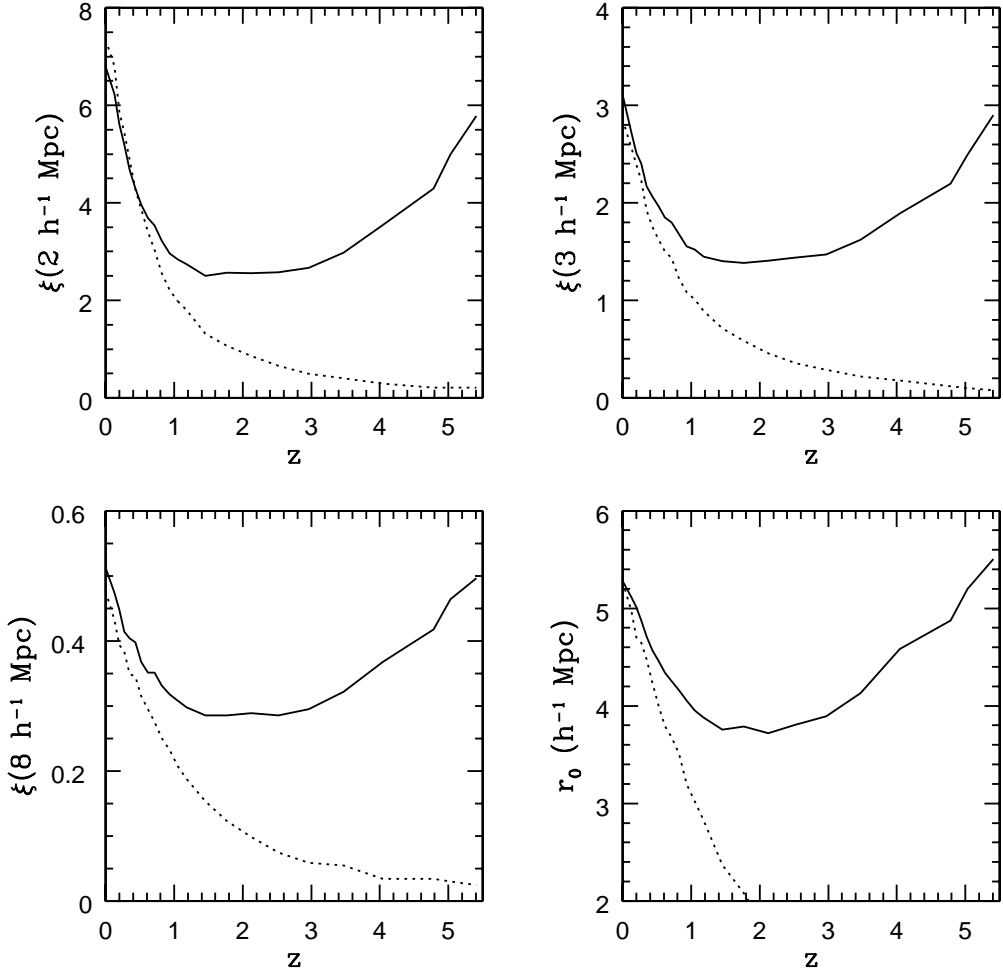}
}
\caption{\label{fig3}
Evolution of clustering the the $\Lambda$CDM model.
In the first 3 panels, the clustering amplitude is plotted against redshift for galaxies  
with rest-frame B-band magnitude brighter than $-19 +5 \log h$ (solid lines)
and for the dark matter (dotted line). Results are shown for $\xi(r)$ evaluated
at $r= 2$, 3 and 8 h$^{-1}$ Mpc$^{-1}$. In the fourth panel, the comoving
correlation length $r_0$ is plotted against redshift both for the
galaxies and for the dark matter.}
\end {figure}

\begin{figure}
\centerline{
\epsfxsize=14cm \epsfbox{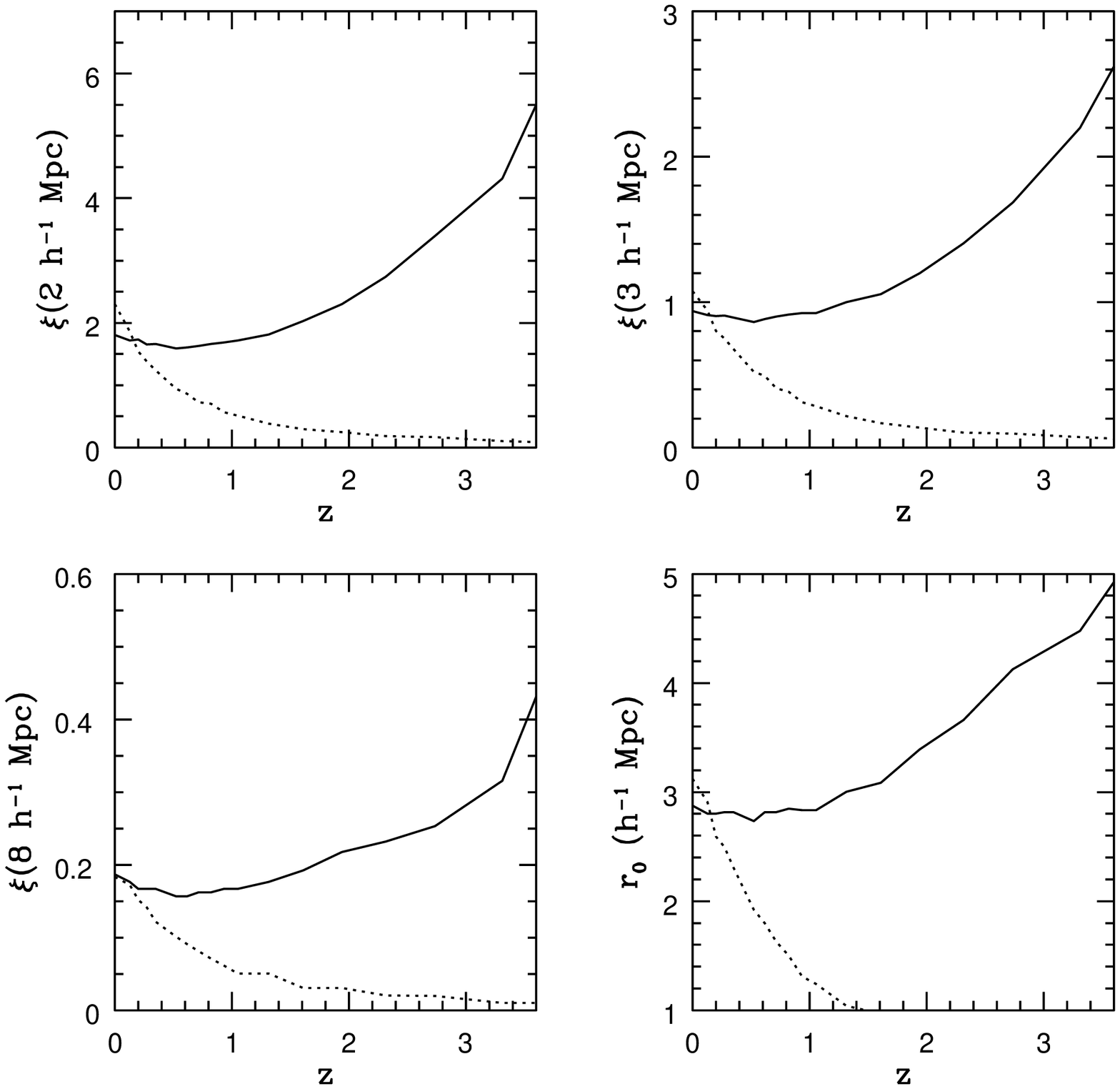}
}
\caption{\label{fig4}
As in figure 3, except for the $\tau$CDM model.} 
\end {figure}

In figure 5, we plot the evolution of the bias $b$, 
defined as the square root of the ratio between the galaxy and the
dark matter correlation functions:
\begin {equation} b(r) = \left(  \frac{\xi_g (r)} {\xi_{m} (r)} \right) ^{1/2}. \end {equation}  
The four lines on the plot show the bias as a function of redshift evaluated at
$r=$ 2, 3, 5 and 8 h$^{-1}$ Mpc. The bias
does not depend on $r$ in either model, except at very high redshifts where $b$
is  somewhat larger on small scales. The bias evolves much more rapidly in the $\tau$CDM
model than in the $\Lambda$CDM model. Galaxies in the $\Lambda$CDM model are unbiased    
tracers of the mass out  to $z \sim 1$. Galaxies in the $\tau$CDM model
have bias values of about 2 at this redshift.

\begin{figure}
\centerline{
\epsfxsize=8cm \epsfbox{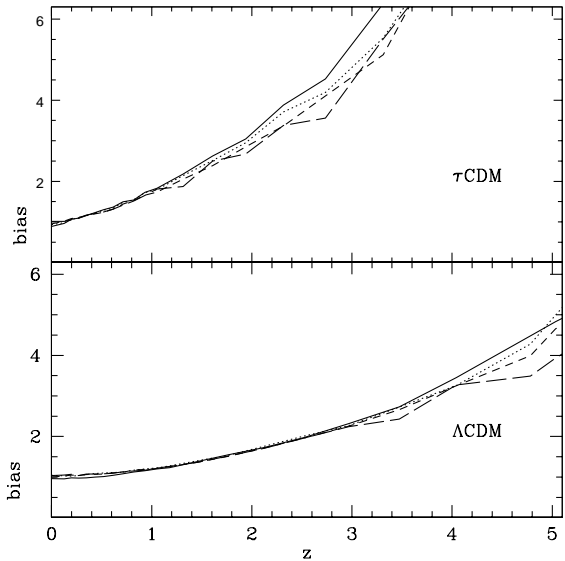}
}
\caption{\label{fig5}
Evolution of the bias for galaxies with rest-frame B-band magnitude brighter
than $-19+5 \log h$ in the $\Lambda$CDM and $\tau$CDM models. Solid, dotted, short-dashed
and long-dashed lines show results evaluated on comoving scales of 2,
3, 5 and 8 h$^{-1}$ Mpc respectively.} 
\end {figure}

\section {Dependence on luminosity, star formation rate and morphological type}
In figures 6 and 7, we demonstrate that the 
clustering evolution depends on the way in which galaxies are selected in the simulations.
The first panel compares the clustering evolution of galaxies selected in the rest-frame B-band
with that of galaxies selected in the rest-frame I-band.
The second panel compares the  clustering evolution of galaxies with $M(B) < -19+5 \log h$
with that of galaxies 1.5 magnitudes brighter.
The third panel shows what happens if galaxies are selected by 
star formation rate rather than by luminosity. The fourth panel shows the clustering of 
early-type galaxies with stellar mass greater than $3 \times 10^{10}
M_{\odot}$ (recall from Paper I that these objects form by mergers of
two galaxies of similar mass and have $M(B)_{bulge}-M(B)_{total} < 1$ mag).

In the $\Lambda$CDM model, we find that the strength of the ``dip'' in clustering
between $z=0$ and $z=1.5$ is sensitive
to sample selection. The dip is weaker for more luminous galaxies and for galaxies
selected in the I-band, but stronger for galaxies selected by star formation rate.
The clustering amplitude of early-type galaxies is stronger than that of the population
as a whole and evolves very little with redshift. In the $\tau$CDM model, clustering
evolution is less sensitive to sample selection. The clustering always remains fixed
out to $z \sim 1$ and then rises steeply at higher redshifts.
Early-type galaxies are also very strongly clustered in this model, particularly at high redshifts.

\begin{figure}
\centerline{
\epsfxsize=14cm \epsfbox{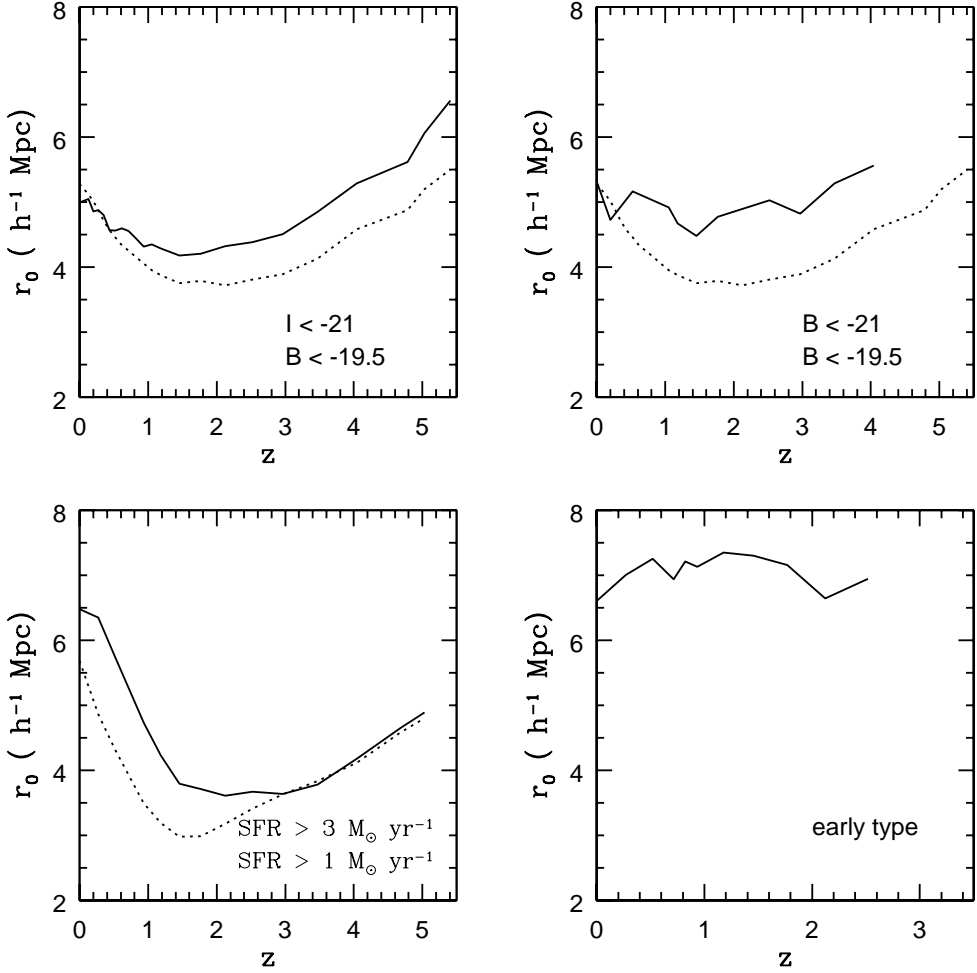}
}
\caption{\label{fig6}
The dependence of clustering evolution on sample selection in the $\Lambda$CDM model.  
The comoving correlation length $r_0$ is plotted as a function of redshift for:
a) $L_*$ galaxies selected in the rest-frame I-band (solid) and
$L_*$ galaxies selected in the rest-frame B-band (dotted);
b) very bright galaxies (solid) and $L_*$ galaxies (dotted);
c) galaxies with star formation rate greater than 3 $M_{\odot}$ yr$^{-1}$
(solid) and 1 $M_{\odot}$ yr$^{-1}$ (dotted);
d) early-type galaxies with stellar masses greater than $3 \times 10^{10} M_{\odot}$.}
\end {figure}

\begin{figure}
\centerline{
\epsfxsize=14cm \epsfbox{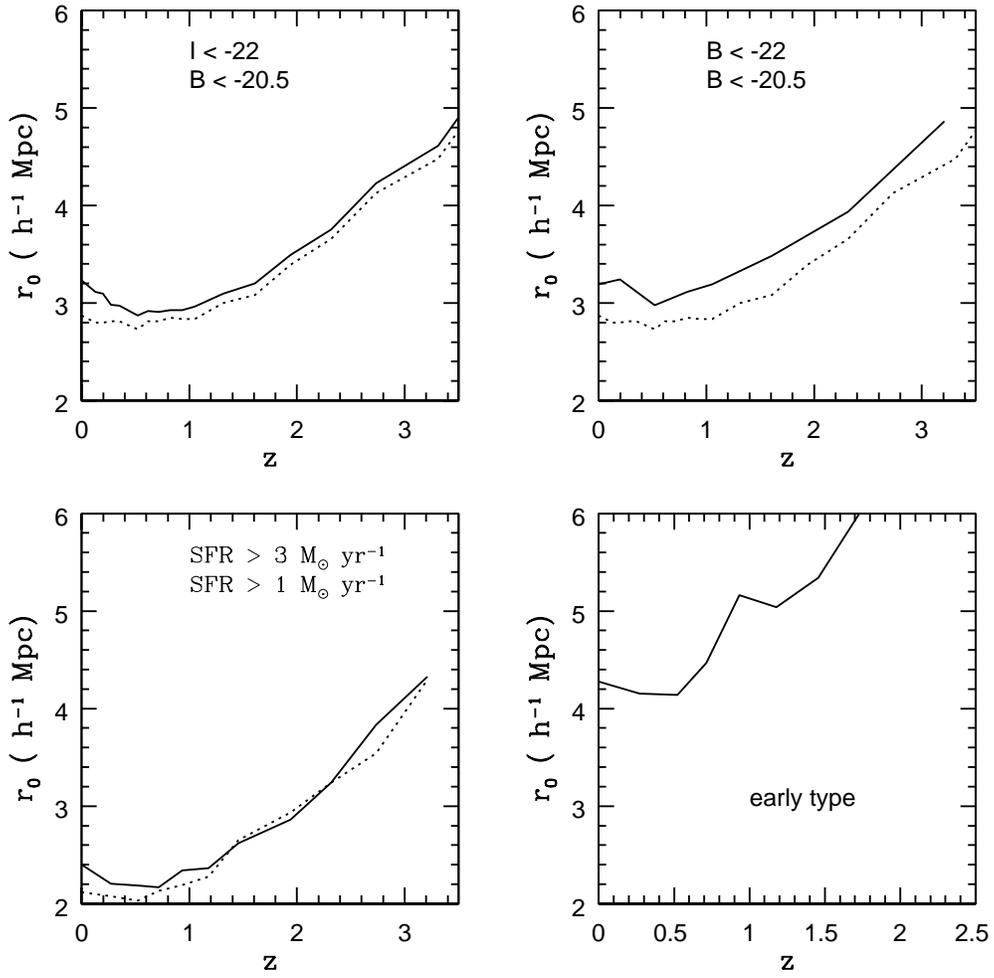}
}
\caption{\label{fig7}
As in figure 6, except for the $\tau$CDM model.} 
\end {figure}

\subsection {What can be learned from these dependences?}
We now  explain {\em why} the evolution of clustering 
depends on sample selection and {\em what }can be learned about galaxy
formation by studying the observed evolution as a function of 
morphological type, luminosity, colour and star formation rate.

For a given cosmology, the clustering amplitude predicted for a sample 
of galaxies depends on the masses
of the dark matter halos they inhabit. The evolution of
clustering depends on how the mass distribution of these halos 
changes with redshift. Additional relevant and observationally accessible
information comes from the variation with redshift of the abundance 
of galaxies in the sample.

As an example, let us suppose that galaxies with fixed star formation rate                         
are found in smaller halos at high redshift than at the present day. We would expect the
clustering amplitude of a SFR-selected sample to show a stronger dip than a sample of galaxies
that tracked halos of the same mass at all redshifts. We would also expect the abundance
of galaxies in a SFR-selected sample to increase more strongly with redshift, because there are many more small
halos than large ones.

As a second example, let us suppose that early type galaxies are found primarily          
in massive halos at all redshifts. As seen in figure 1, these galaxies should not 
exhibit any dip in clustering
and their abundances should decrease strongly at high redshifts
because massive halos are rare objects at early times.

These points  are illustrated in detail in figure 8, where we plot the 
evolution of the median halo mass and the comoving number density of galaxies in samples selected in different ways
from the  $\Lambda$CDM  simulation. 
The top 3 panels show results for
galaxies selected according to  rest-frame B-magnitude, rest-frame I-magnitude and star formation
rate. At $z=0$, all three galaxy  samples have the same abundance and occur in halos of roughly
the same mass. Galaxies selected according to star formation rate move to smaller halos
at higher redshift. This effect is simply a result of the parametrization of star formation in our models.
Following Kennicutt (1998), we have adopted a star formation law of the form $\dot{M}_* = \alpha M_{cold}/t_{dyn}$,
where $M_{cold}$ is the mass of cold gas in the galaxy and $t_{dyn}$ is the dynamical time
of the galaxy. Since $t_{dyn}$ decreases at higher redshifts, the star formation rates are higher in halos of
the same cold gas content. Galaxies selected in the B-band exhibit a weaker trend towards low-mass halos. 
In the case of the I-band selection, galaxies trace halos of roughly the same mass
at all redshifts below 2. This is because the I-band magnitude of a galaxy is  a measure
of its total stellar mass, rather than its instantaneous star formation rate.
We thus conclude that SFR-selected samples show the strongest dip in clustering in figure 6 because
this selection procedure favours galaxies in lower mass halos at high redshift.
Note that galaxies in the SFR-selected samples also exhibit                 
the strongest increase in abundance from $z=0$ to $z=1.5$.

The bottom two panels in figure 8 show that very bright galaxies and  early-type
galaxies in the simulation are found in halos with masses $\sim 10^{13} M_{\odot}$.
As seen in figure 1, the clustering of these objects evolves very little from $z=0$ to $z=1$.

\begin{figure}
\centerline{
\epsfxsize=14cm \epsfbox{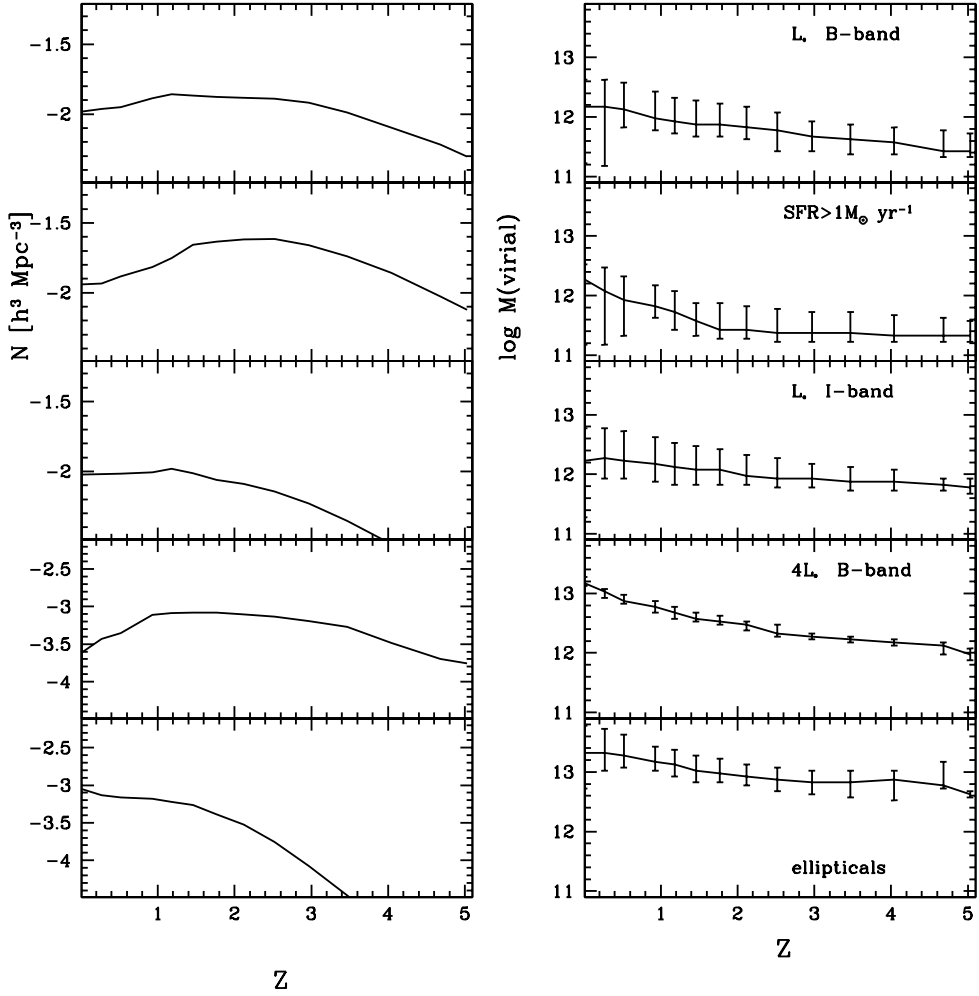}
}
\caption{\label{fig8}
The evolution of the comoving number density (left column) and the median halo mass (right column)
of galaxies selected from the $\Lambda$CDM simulation. Error bars
indicate the upper and lower quartiles of the halo mass
distributions. Results are shown for the 
selection criteria described in the caption to figure 6. }
\end {figure}

In figure 9, we show the evolution of galaxy abundances and median halo masses for samples           
selected in the same way from the $\tau$CDM simulation. The results
are qualitatively similar to
those found for $\Lambda$CDM. $L_*$ galaxies occur in halos of roughly the same masses
($\sim 10^{12} M_{\odot}$) in both models. The reason why no dip is seen in the $\tau$CDM model is because          
halos of these masses are more strongly biased at $z=1$ than in the $\Lambda$CDM model and the  
decrease in halo mass with redshift for the SFR-selected sample is less pronounced.
Note also that the redshift at which the abundance curves peak 
is higher for $\Lambda$CDM  than for  $\tau$CDM.
In the $\Lambda$CDM simulation, the abundance of early-type galaxies only decreases
substantially at redshifts greater than 1.5, whereas in the $\tau$CDM simulation, the abundance
of ellipticals has already declined by
a factor of 3 by $z=1$.

\begin{figure}
\centerline{
\epsfxsize=14cm \epsfbox{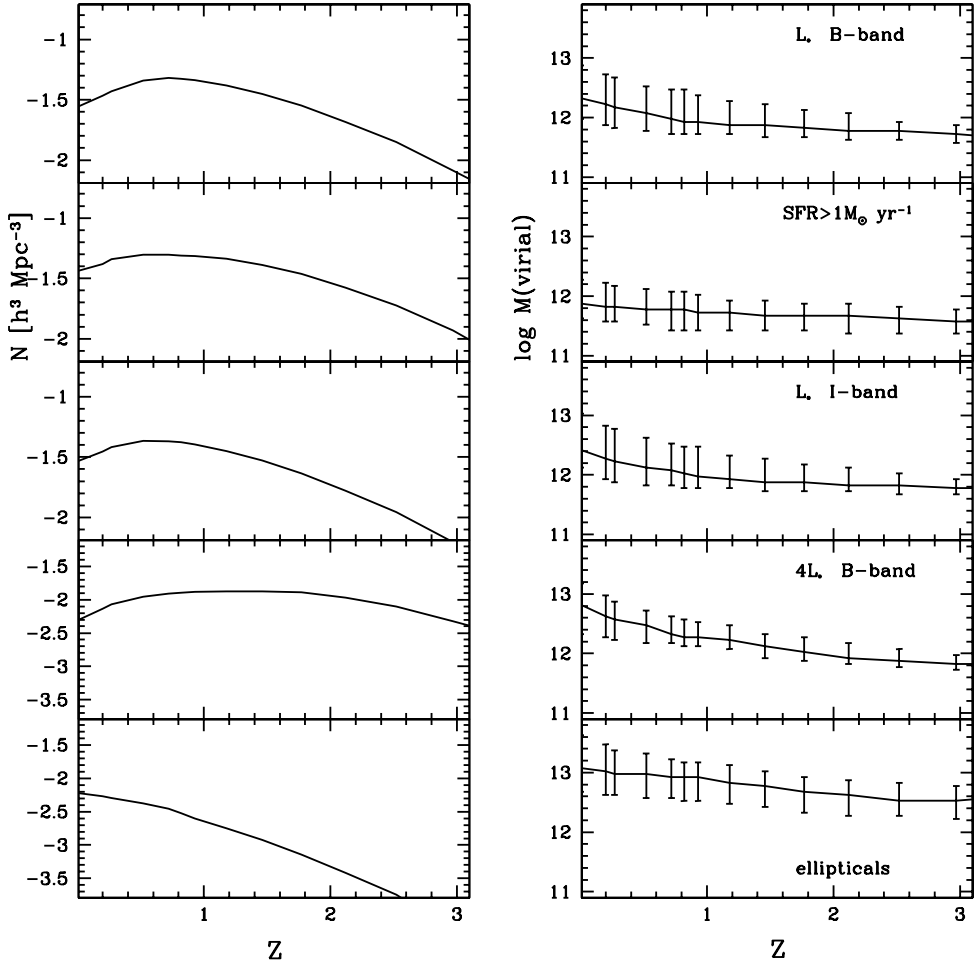}
}
\caption{\label{fig9}
As in figure 8, except for the $\tau$CDM simulation.}
\end {figure}

\section {Evolution of the slope of $\xi(r)$} 
Figure 10 shows the evolution of the slope $\gamma$ of the two-point correlation
function for galaxies with rest-frame B-magnitudes brighter than $-19+5 \log h$ in the
$\Lambda$CDM and $\tau$CDM  models. We have fit a power-law to $\xi(r)$ over three different ranges
in scale: $r= 1 -5$ h$^{-1}$ Mpc, $r= 5-10$ h$^{-1}$ Mpc and $r=1-10$ h$^{-1}$ Mpc.

In the $\Lambda$CDM model, the evolution of the slope is stronger on small scales. 
Over the range $1-5$ h$^{-1}$ Mpc, 
$\gamma$ evolves from $-2$ at $z=0$ to $-1.5$ at $z=1$. On large scales, $\gamma$ remains
approximately constant. Over the range $1-10$ h$^{-1}$ Mpc, $\gamma$ evolves
from $-1.85$ at $z=0$ to $-1.6$ at $z=1$ and then remains constant.
These results appear to be in qualitative agreement with the observations (Postman et al 1998).
These au thors find no dependence of $\gamma$ on 
magnitude for the bright ($I < 21$) galaxies in their
survey. At fainter magnitudes $\gamma$ flattens, reaching a value of $-1.6$ at I=22.5. They also find that
the flattening is stronger on smaller angular scales. Neuschaeffer \&  Windhorst (1995)
find similar results from an independent survey carried out at a different
wavelength.                 
In the $\tau$CDM model, there is very little change in the slope with
redshift on any scale.

\begin{figure}
\centerline{
\epsfxsize=11cm \epsfbox{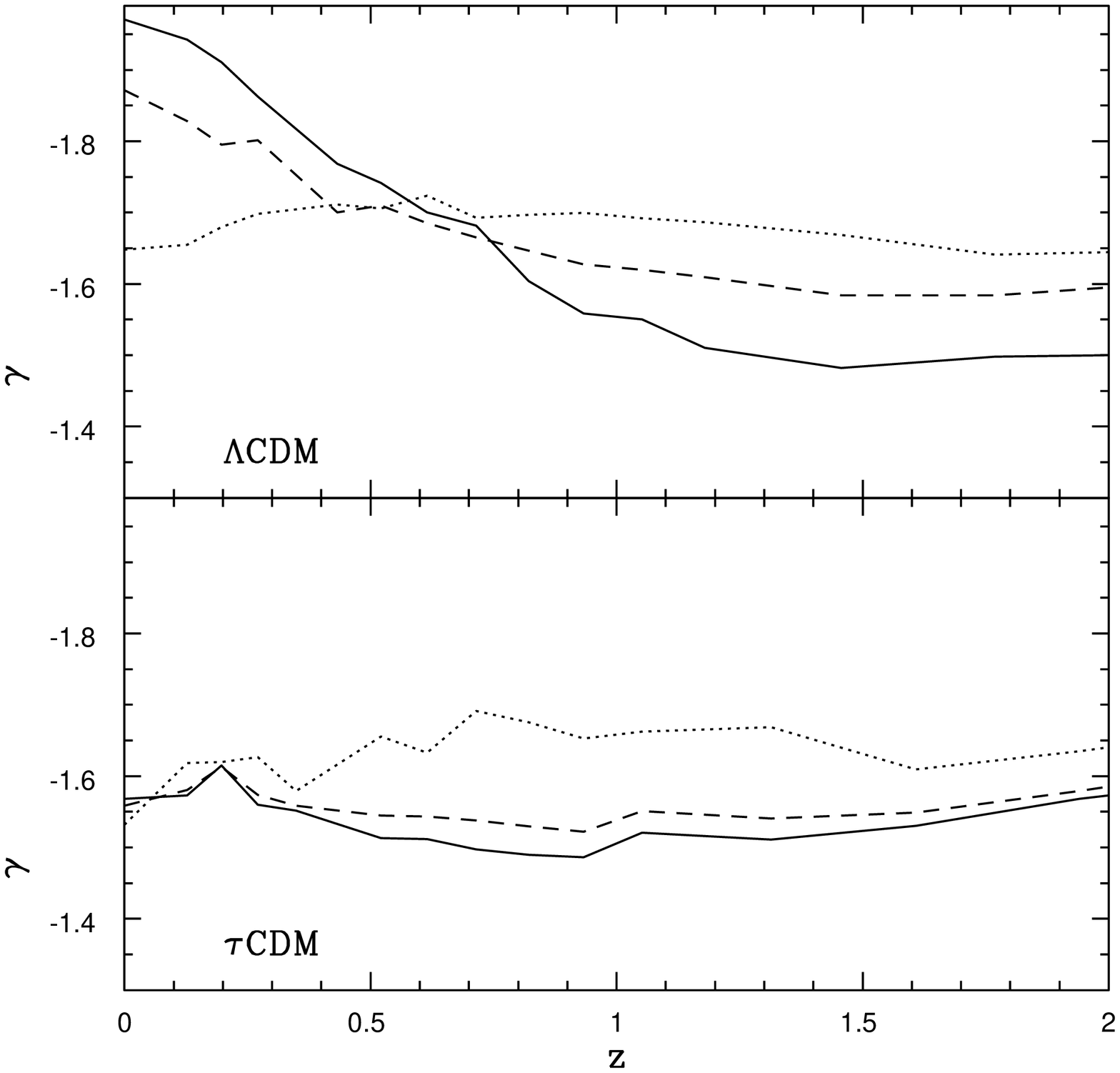}
}
\caption{\label{fig10}
The evolution of the slope of the correlation function of galaxies with $M(B) < -19 +5 \log h$
in the $\Lambda$CDM and $\tau$CDM simulations. The solid line is the result of a fit to $\xi(r)$ over scales between
1 and 5 h$^{-1}$ Mpc, the dotted line is for scales between 5
and 10 h$^{-1}$ Mpc and the dashed line for scales between 
1 and 10 h$^{-1}$ Mpc.}
\end {figure}

\section {Evolution of pairwise peculiar velocities}
Estimates of dynamical quantities such as $\Omega$ or cluster M/L
ratios from galaxy data require knowledge not only of the spatial bias
in the galaxy distribution, but also of any possible bias in the
kinematics of the galaxies relative to those of the dark matter.
In Paper I we explored this `` velocity bias'' for our $z=0$ models
using pairwise velocity statistics, and in Paper III (Diaferio et al
1998)  we will do the same using group and cluster velocity
dispersions. Here we briefly explore the predicted evolution of
velocity bias using pairwise statistics.
The thin solid lines in figure 11 show the redshift evolution of the pairwise peculiar velocity 
dispersion $\sigma_{12}$ evaluated at relative separations $r=$ 0.5, 1 and 
2 h$^{-1}$ Mpc (co-moving units) for
galaxies with rest-frame B-band magnitudes less than $-19+5\log h$ in the $\Lambda$CDM and
$\tau$CDM simulations. The thick solid lines show the evolution of $\sigma_{12}$ 
for the dark matter. In order to compare the {\em relative change} in $\sigma_{12}$
as a function of redshift in the two models, we scale the results by dividing by the value
of $\sigma_{12}^{gal}$ at z=0. As shown in figure 13 of Paper I, $\sigma_{12}^{gal} \simeq
800$ km s$^{-1}$ ($r=1$ h$^{-1}$ Mpc) in both the $\Lambda$CDM and $\tau$CDM models at the present day.

The pairwise peculiar velocities of the galaxies follow those of the dark matter
quite closely in both models. The galaxy velocities are 10-40\% lower than those of the
dark matter at $z=0$. (Note that the ``antibias'' in galaxy peculiar velocities at $z=0$
is stronger than that shown in figure 13 of Paper I, because the models presented in this
paper do not include dust extinction. Dust reduces the contribution of star-forming
field galaxies in a B-selected sample, but has little effect on  early-type galaxies
in rich groups and clusters. Models that include dust extinction thus
give values of $\sigma_{12}^{gal}$ that are 10-25 \% larger ). The difference between the
galaxy and dark matter peculiar velocities decreases at higher
redshift. In contrast to the spatial distributions, galaxy peculiar
velocities in our models are never very  strongly biased.
In the $\tau$CDM model, there is nearly a factor 2 decrease in $\sigma_{12}^{gal}$ 
from z=0 to z=0.5.
In the $\Lambda$CDM models, $\sigma_{12}^{gal}$ remains roughly constant out
to $z=0.5$, before decreasing at higher redshift.

\begin{figure*}
\vbox to11.cm{\rule{0pt}{10.cm}}
\includegraphics{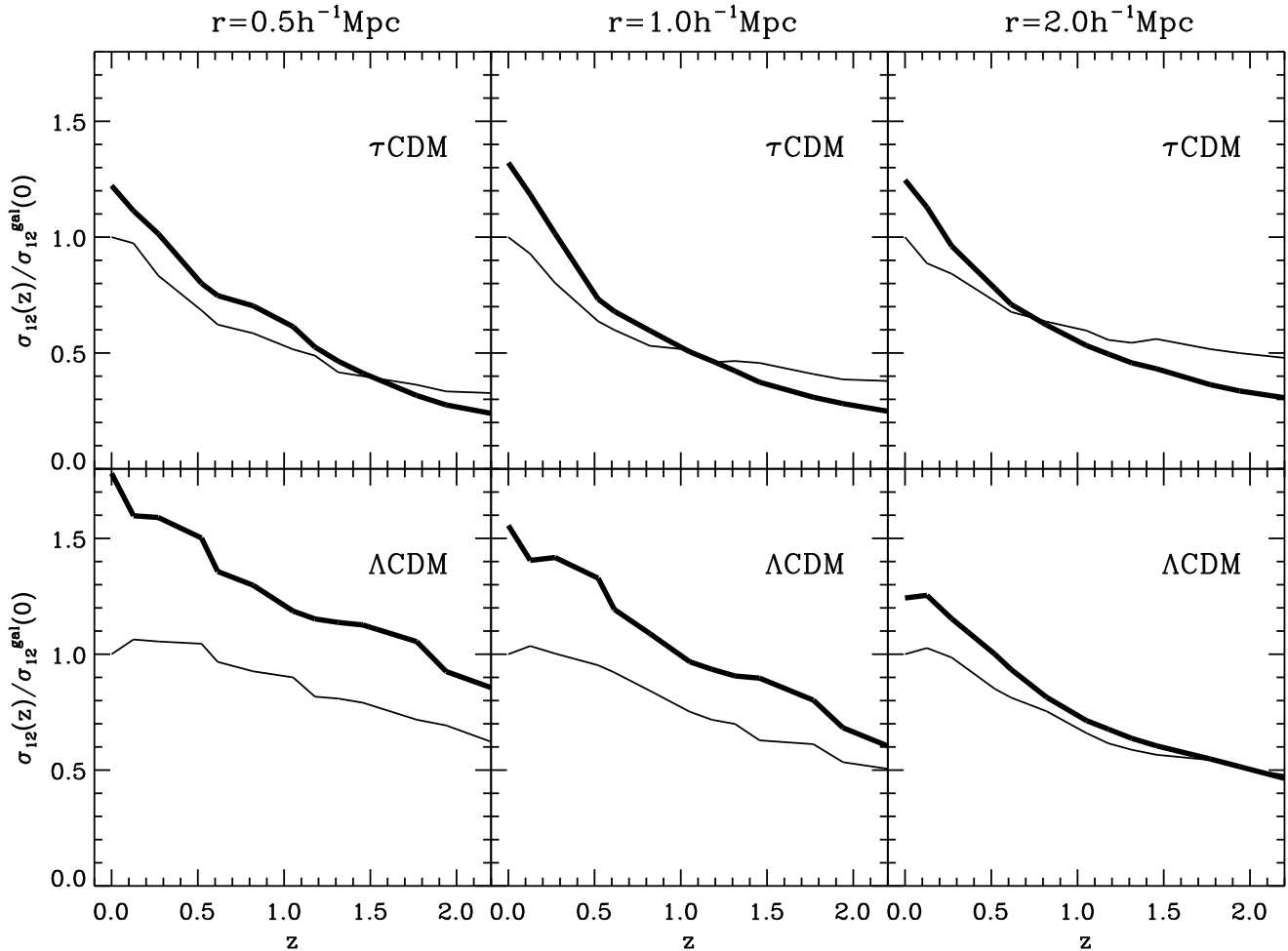}
\caption{
Redshift evolution of the  pairwise velocity dispersion
$\sigma_{12}$ at  relative comoving separations of 0.5, 1 and 2 h$^{-1}$ Mpc for galaxies
with rest-frame B-magnitudes brighter than $-19+5\log h$ (thin lines)
and for dark matter (thick lines) in the $\Lambda$CDM and $\tau$CDM simulations.
The results are scaled by dividing by the value of $\sigma^{gal}_{12}$ at
$z=0$.}
\label{fig:11}
\end{figure*}

\section {Discussion and Conclusions}
In a hierarchical Universe, the evolution of galaxy clustering depends on the following:
\begin {enumerate}
\item Cosmological parameters, such as $\Omega$, $\Lambda$ and $\sigma_8$, because these determine
the rate at which structure grows and the epoch at which halos of
given mass change from
being rare objects, and thus  biased tracers of the dark matter distribution, to being
``typical'' objects with clustering properties similar to that of the mass.
\item  The relationship between the mass of a dark matter halo and the
properties of the galaxies that form within it. Note that
this relationship depends {\em only} on halo mass and is independent
of the environment in which the halo finds itself (Lemson \& Kauffmann
1998).
\item The evolution of the galaxy population with redshift (and hence the evolution of the
  relationship between galaxy properties and halo mass).                     
\end {enumerate}

In this paper, we illustrate how differences in clustering evolution between
galaxies of differing luminosity, colour, morphological type and 
star formation rate may help constrain
galaxy formation models, and perhaps even cosmological parameters.

One interesting diagnostic that we highlight is the ``dip'' in
correlation amplitude  observed between $z=0$ and $z=1$. We show that
this dip occurs naturally in a $\Lambda$CDM model, where structure
forms early and halos with masses in the range $10^{11} - 10^{12} M_{\odot}$,
which contain galaxies of intermediate luminosities, are unbiased tracers of the mass            
over this redshift range. In the $\tau$CDM model, bias evolves rapidly,
and the clustering amplitude of $L_*$ galaxies remains constant  
from $z=0$ to $z=1$. Although it might be possible to ``force'' a dip in clustering
in the $\tau$CDM model by requiring that $L_*$ galaxies form in less massive halos,         
it would be difficult to come up with a physically-motivated scheme for
doing this that would not simultaneously produce too many bright galaxies.

We also show that the strength of the dip in the $\Lambda$CDM
model is sensitive to sample selection. If galaxies are selected
according to star formation rate rather than B-band luminosity, objects
in low mass halos contribute more to the  clustering signal at high
redshifts and the dip is stronger.
If galaxies are selected in the red rather than the blue, the
dip is reduced. Very luminous galaxies
and massive early-type galaxies exhibit no dip in clustering between $z=0$ and
$z=1$ because they occur in high mass halos ($10^{13}-10^{14} M_{\odot}$)
that are already biased at $z=0$ and become substantially more biased
at high redshift.

The predictions presented in this paper should be viewed as illustrative rather
than quantitative. As discussed in Paper I, the precise relation
between the mass of a halo and the properties of the galaxies that
form within it depends strongly on the adopted recipes
for star formation and feedback; these are  very uncertain.
The exciting prospect is that future observations of galaxy clustering at high redshift
will place strong {\em empirical} constraints
on these processes.

\vspace{0.8cm}

\large
{\bf Acknowledgments}\\
\normalsize
The simulations in this paper were carried out at the Computer Center of the Max-Planck Society 
in Garching and at the EPPC in Edinburgh. Codes were kindly made available by the Virgo Consortium.
We especially thank Adrian  Jenkins and Frazer Pearce for help in carrying them out.
We are also grateful to John Peacock for useful discussions.                            
A.D. is a Marie Curie Fellow and holds grant ERBFMBICT-960695 of the Training and Mobility of
Researchers program financed by the EC.

\pagebreak
\Large
\begin {center} {\bf References} \\
\end {center}
\normalsize
\parindent -7mm
\parskip 3mm

Adelberger, K.L., Steidel, C.C., Giavalisco, M., Dickinson, M.,
Pettini, M. \& Kellogg, M., 1998, astro-ph/9804236    

Bagla, J.S., 1998a, MNRAS, 297, 251

Bagla, J.S., 1998b, astro-ph/9711081

Baugh, C.M., Cole, S., Frenk, C.S. \& Lacey, C.G., 1998, ApJ, 498, 504

Brainerd, T.G. \& Villumsen, J.V., 1994, ApJ, 431, 477

Carlberg, R.G., Yee, H.K.C., Morris, S.L., Lin, H., Sawicki, M., Wirth, G., Patton, D., Shepherd, C.W. et al,
1998, astro-ph/9805131

Coles, P., Lucchin, F., Matarrese, S. \& Moscardini, L., 1998, astro-ph/9803197

Connolly, A.J., Csabai, I., Szalay, A.S., Koo, D.C., Kron, R.G. \& Munn, J.A., 1995, AJ, 110, 2655

Giavalisco, M., Steidel, C.C., Adelberger, K.L., Dickinson, M.,
Pettini, M. \& Kellogg, M., 1998, ApJ, 503, 543

Governato, F., Baugh, C.M., Frenk, C.S., Cole, S. \& Lacey, C.G., 1998, Nature, 392, 359

Hamilton, A.J., Kumar, P., Lu, E. \& Matthews, S.A., 1991, ApJ, 374, L1

Jain, B., Mo, H.J. \& White, S.D.M., 1995, MNRAS, 276, L25

Jenkins, A., Frenk, C.S., Pearce, F.R., Thomas, P.A., Colberg, J.M., White, S.D.M.,
Couchman, H.M.P., Peacock, J.A., Efstathiou, G. \& Nelson, A.H., 1998, ApJ, 499, 20 

Jing, Y.P. \& Suto, Y., 1998, ApJ, 494, 5

Kauffmann, G., Colberg, J.M., Diaferio, A. \& White, S.D.M., 1998, astro-ph/9805283

Kaiser, N., 1984, ApJ, 284, 9

Kennicutt, R.C., 1998, ApJ, 498, 541

Le F\`{e}vre, O., Hudon, D., Lilly, S.J., Crampton, D., Hammer, F. \& Tresse, L., 1996, ApJ, 461, 534

Lemson, G. \& Kauffmann, G., 1998, astro-ph/9710125

Ma, C.P., 1998, astro-ph/9808130

Matarrese, S., Coles, P., Lucchin, F. \& Moscardini, L., 1997, MNRAS, 286, 115

Mo, H.J. \& Fukugita, M., 1996, APJ, 467, L9

Mo, H.J., Mao, S. \& White, S.D.M., 1998, astro-ph/9807341

Mo, H.J. \& White, S.D.M., 1996, MNRAS, 282, 347

Neuschaeffer, L.W. \& Windhorst, R.A., 1995, ApJ, 439, 14

Peacock, J.A. \& Dodds, S.J., 1994, MNRAS, 267, 1020

Postman, M., Lauer, T.R., Szapudi, I. \& Oegerle, W., astro-ph/9804141

Roukema, B.F., Peterson, B.A., Quinn, P.J. \& Rocca-Volmerange, B., 1997, MNRAS, 292, 835

Somerville, R.S., Primack, J.R. \& Faber, S.M., astro-ph/9806228

Steidel, C.C., Adelberger, K.L., Dickinson, M., Giavalisco, M., Pettini,M. \& Kellogg,M., 1998a, ApJ, 492, 428 

Steidel, C.C., Adelberger, K.L., Giavalisco, M., Dickinson, M.E., Pettini,M. \& Kellogg,M., 1998b, astro-ph/9805267

Vogt, N.P., Forbes, D.A., Phillips, A.C., Gronwall, C., Faber, S.M., Illingworth, G.D. \& Koo, D.C., 1996, ApJ,
465,15

Wechsler, R.H., Gross, M.A.K., Primack, J.R., Blumenthal, G.R. \&
Dekel, A., 1998,  astro-ph/9712141

\end {document}